\newcommand{\CLASSINPUTtoptextmargin}{19mm}
\newcommand{\CLASSINPUTbottomtextmargin}{25.4mm}
\newcommand{\CLASSINPUTinnersidemargin}{16mm}
\newcommand{\CLASSINPUToutersidemargin}{16mm}

\documentclass[conference,10pt,a4paper]{IEEEtran}
\IEEEoverridecommandlockouts
\usepackage{amsmath,amssymb,amsfonts}
\usepackage{algorithmic}
\usepackage{graphicx}
\usepackage{textcomp}
\usepackage{xcolor}
\usepackage{biblatex}
\def\BibTeX{{\rm B\kern-.05em{\sc i\kern-.025em b}\kern-.08em
    T\kern-.1667em\lower.7ex\hbox{E}\kern-.125emX}} 
\renewcommand\IEEEkeywordsname{Keywords} 
\newcommand{\IEEEtitle}[1]{\title{\vspace{-6.5mm}#1}}
\newcommand{\IEEEand}{\\\vspace{-12mm}\and}
\DeclareUnicodeCharacter{01FF}{\ocircumflex}
\usepackage{hyperref}
\begin{document}

% title and acknowledgement
\IEEEtitle{PERU MINING: ANALYSIS AND FORECAST OF MINING PRODUCTION IN PERU USING TIME SERIES AND DATA SCIENCE TECHNIQUES\\
%\IEEEThanks{Identify applicable funding agency here. If none, delete this.Identify applicable funding agency here. If none, delete this.Identify applicable funding agency here. If none, delete this.Identify applicable funding agency here. If none, delete this.Identify applicable funding agency here. If none, delete this.Identify applicable funding agency here. If none, delete this.Identify applicable funding agency here. If none, delete this.Identify applicable funding agency here. If none, delete this.}
}

% authors and affiliations
\author{

\IEEEauthorblockN{Aycaya-Paco Yhack Bryan}
\IEEEauthorblockA{\textit{Facultad de Ingeniería Estadística e Informática} \\
\textit{Universidad Nacional del Altiplano}\\
Puno, Perú\\
Email: yaycaya@est.unap.edu.pe}\\
\IEEEauthorblockN{Torres-Cruz Fred}
\IEEEauthorblockA{\textit{Facultad de Ingeniería Estadística e Informática} \\
\textit{Universidad Nacional del Altiplano}\\
Puno, Perú \\
Email: ftorres@unap.edu.pe}\\

\IEEEand % new column

\IEEEauthorblockN{Vilca-Mamani Lindell Dennis}
\IEEEauthorblockA{\textit{Facultad de Ingeniería Estadística e Informática} \\
\textit{Universidad Nacional del Altiplano}\\
Puno, Perú \\
Email: livilca@est.unap.edu.pe}\\

%\IEEEauthorblockN{3\textsuperscript{th} Torres-Cruz Fred}
%\IEEEauthorblockA{\textit{Facultad de Ingeniería Estadística e Informática} \\
%\textit{Universidad Nacional del Altiplano}\\
%Puno, Perú \\
%Email: ftorres@unap.edu.pe}\\

\IEEEand % new column

}

\maketitle

\begin{abstract}
Peruvian mining plays a crucial role in the country's economy, being one of the main producers and exporters of minerals worldwide. In this project, an application was developed in RStudio that utilizes statistical analysis and time series modeling techniques to understand and forecast mineral extraction in different departments of Peru. The application includes an interactive map that allows users to explore Peruvian geography and obtain detailed statistics by clicking on each department. Additionally, bar charts, pie charts, and frequency polygons were implemented to visualize and analyze the data. Using the ARIMA model, predictions were made on the future extraction of minerals, enabling informed decision-making in planning and resource management within the mining sector. The application provides an interactive and accessible tool to explore the Peruvian mining industry, comprehend trends, and make accurate forecasts. These predictions for 2027 in total annual production are as follows: Copper = 2,694,957 MT, Gold = 72,817.47 kg Fine, Zinc = 1,369,649 MT, Silver = 3,083,036 MT, Lead = 255,443 MT, Iron = 15,776,609 MT, Tin = 29,542 MT, Molybdenum = 35,044.66 MT, and Cadmium = 724 MT. These predictions, based on historical data, provide valuable information for strategic decision-making and contribute to the sustainable development of the mining industry in Peru.\\
\end{abstract}


\section{Introduction}
Peruvian mining has played a fundamental role in the country's economy, contributing significantly to its growth and development \cite{minem}. In this project, we focus on the analysis and optimization of the Peruvian mining industry using advanced statistical analysis tools. Our objective is to gain a better understanding of the functioning of this industry and maximize its efficiency.
Peru is globally recognized for its abundance of mineral resources such as copper, gold, silver, and zinc. These resources have attracted foreign investments and positioned the country as one of the main producers and exporters of minerals. Mining has generated employment, driven the growth of local communities, and contributed to the development of infrastructure in mining regions \cite{snmpe}.

However, the mining industry also faces significant challenges. Efficient resource management, reducing environmental impact, and sustainability are key aspects that need to be addressed to ensure responsible mining development \cite{cardenas2016vulnerabilidad}. In this context, statistical analysis presents itself as a valuable tool to understand trends, patterns, and factors influencing mining production.
This project is based on previous research and innovative approaches in statistical analysis applied to Peruvian mining \cite{Sanchez2017Analisis}. Studies have shown how statistical analysis can help identify improvement opportunities in mining production and optimize operational processes \cite{zarate2020industria}. Additionally, the report from Peru's Ministry of Energy and Mines, provides an overview of the current mining situation in the country and highlights the importance of efficient natural resource management.

In this context, we have developed the web application MineAnalytica using Shiny/R Studio. MineAnalytica is a powerful tool that allows visualization and analysis of mining production data using time series techniques and ARIMA and state space models \cite{forecast_page}. Its aim is to predict trends and provide future estimates of metallic mining production in Perú.

MineAnalytica is built on a solid foundation of references and reliable sources. These include the Geological, Mining, and Metallurgical Institute (INGEMMET) \cite{ingemmet}, which provides up-to-date geological and technical data on Peruvian mining, and the Central Reserve Bank of Peru \cite{bcrp}, which offers economic information and relevant statistics for the mining sector. 

Time series techniques, supported by studies \cite{szablowski2002mining}, enable users to make accurate forecasts about mining production and anticipate possible future scenarios. This facilitates strategic decision-making and contributes to more efficient management of mineral resources \cite{lalama2017}.

In summary, our project focuses on the analysis, visualization, and prediction of mining production in Peru. By using MineAnalytica as the main tool, supported by solid research and collaborations with leading institutions in the mining sector, we aim to provide users with an effective tool for informed and strategic decision-making. By facilitating access to updated data and advanced analysis tools, we contribute to the sustainable and responsible development of the mining industry in the country.

\section{Methodology}
In this section, we describe the methodology used to conduct our study on mining production in Peru and the development of the web application MineAnalytica.

\subsection{Data Collection}

We obtained the mining production data for Peru from the official portal of MINEM \cite{minem}. We selected two sets of data: monthly mining production from 2020 until the end of 2022 and annual mining production data from 1900. These datasets allow us to analyze both recent patterns and long-term trends in the Peruvian mining industry.

\begin{table}[htbp]
\centering
\caption{Distribution of the dataset}
\begin{tabular}{ccc}
\hline
\textbf{Variable} & \textbf{Total} & \textbf{Type} \\
\hline
Mineral & 2151 & chr \\
Unidad de medida & 2151 & chr \\
Etapa & 2151 & chr \\
Proceso & 2151 & chr \\
Estrato & 2151 & chr \\
Titular & 2151 & chr \\
Departamento & 2151 & chr \\
Año & 2151 & num \\
Enero & 2151 & num \\
Febrero & 2151 & num \\
Marzo & 2151 & num \\
Abril & 2151 & num \\
Mayo & 2151 & num \\
Junio & 2151 & num \\
Julio & 2151 & num \\
Agosto & 2151 & num \\
Septiembre & 2151 & num \\
Octubre & 2151 & num \\
Noviembre & 2151 & num \\
Diciembre & 2151 & num \\
Total & 2151 & num \\
\hline
\end{tabular}
\end{table}

\begin{table}[htbp]
\centering
\caption{Mineral Production 1980-2022}
\begin{tabular}{ccc}
\hline
\textbf{Variable} & \textbf{Total} & \textbf{Type} \\
\hline
AÑO & 43 & num \\
COBRE(TMF) & 43 & num \\
ORO(KG) & 43 & num \\
ZINC(TMF) & 43 & num \\
PLATA(TMF) & 43 & num \\
PLOMO(TMF) & 43 & num \\
ESTAÑO(TMF) & 43 & num \\
MOLIBDENO(TMF) & 43 & num \\
CADMIO(TMF) & 43 & num \\
\hline
\end{tabular}
\end{table}

\subsection{Data Analysis and Processing}

We conducted a data processing and cleaning process to ensure the quality and consistency of the information \cite{chu2016data}. During this stage, we applied techniques such as variable name correction, removal of empty or null data, and imputation of missing values using the k-Nearest Neighbors (k-NN) \cite{garcia2018tratamiento} algorithm to obtain accurate estimations.

Additionally, we also applied techniques to adjust the monthly data and appropriately adapt it for use in the ARIMA model. This involved performing transformations or adjustments to the frequency of the monthly data to meet the requirements of the model. In this way, we could fully leverage the potential of ARIMA models in our time series analysis.

\begin{table}[htbp]
  \centering
  \caption{Data structure}
  \begin{tabular}{cc}
  \hline
\textbf{Variable} & \textbf{Composition} \\
  \hline
  Mineral & Arsénico, cobre, oro, plata, etc. \\
  Unidad de medida & Gr. finos, Kg. finos y TMF \\
  Etapa & Concentración, fundición y refinación \\
  Proceso & Flotación, gravimetría y lixiviación \\
  Estrato & Pequeño productor y régimen general \\
  Departamento & Lima, Puno, Cusco, Arequipa, etc. \\
  Año & Represents year of production \\
  Meses (Ene-Dic) & Represents the quantity extracted each month \\
  Total & Represents total monthly extraction sum \\
  \hline
\end{tabular}
\end{table}

\subsection{Time Series Models}

In this stage, we used time series models to analyze and predict mining production in Peru. Time series models allow us to capture and model temporal patterns and trends present in the data \cite{cryer1986time}.
For our study, we implemented two types of time series models: ARIMA (autoregressive integrated moving average) and state space.

\subsubsection{ARIMA Model}
This model consists of three main components: autoregressive (AR), integrated (I), and moving average (MA) \cite{newbold1983arima}.

\textbf{Autoregressive (AR) Component:} The AR component refers to the linear dependence of a current observation on past values of the time series. It is represented as AR(p), where "p" is the order of the autoregressive component. The general formula for the AR(p) component is:
\[ y(t) = c + \varphi_1 \cdot y(t-1) + \varphi_2 \cdot y(t-2) + \ldots + \varphi_p \cdot y(t-p) + \varepsilon(t) \]
where:
\begin{itemize}
  \item $y(t)$ is the current value of the time series.
  \item $c$ is a constant.
  \item $\varphi_1, \varphi_2, \ldots, \varphi_p$ are the autoregressive coefficients.
  \item $\varepsilon(t)$ is a random error term.
\end{itemize}

\textbf{Moving Average (MA) Component:} The MA component represents the linear dependence of a current observation on past error terms of the time series. It is represented as MA(q), where "q" is the order of the moving average component. The general formula for the MA(q) component is:
\[ y(t) = c + \theta_1 \cdot \varepsilon(t-1) + \theta_2 \cdot \varepsilon(t-2) + \ldots + \theta_q \cdot \varepsilon(t-q) + \varepsilon(t) \]
where:
\begin{itemize}
  \item $y(t)$ is the current value of the time series.
  \item $c$ is a constant.
  \item $\theta_1, \theta_2, \ldots, \theta_q$ are the moving average coefficients.
  \item $\varepsilon(t)$ is a random error term.
\end{itemize}

\textbf{Integration (I):} Integration is used to achieve stationarity in the time series. If the series is not stationary, first-order differencing ($d = 1$) or higher-order differencing can be applied to make it stationary. The general formula for differencing is:
\[ \Delta y(t) = y(t) - y(t-1) \]
where:
\begin{itemize}
  \item $\Delta y(t)$ is the difference between the current value and the previous value of the time series.
\end{itemize}

The ARIMA model combines the linear dependence on past values (AR), linear dependence on past errors (MA), and differencing to model and forecast time series. The parameters $p$, $d$, and $q$ are chosen to fit the model to the data.
\subsubsection{State Space Model}
This model is another approach to analyze and predict time series. It consists of two main components: the state equation and the observation equation \cite{aoki2013state}.

\textbf{State Equation:} The state equation describes how the hidden state of the system evolves over time. It is represented as a linear relationship between the state at the current time and the state at the previous time, with a transition matrix. The general formula for the state equation is:
\[ x(t) = A \cdot x(t-1) + B \cdot u(t) + w(t) \]
where:
\begin{itemize}
  \item $x(t)$ is the state at time $t$.
  \item $A$ is the state transition matrix.
  \item $B$ is the control matrix representing the influence of an external control signal $u(t)$ on the state.
  \item $u(t)$ is the control signal at time $t$.
  \item $w(t)$ is the process noise.
\end{itemize}

\textbf{Observation Equation:} The observation equation relates the hidden state of the system to the measurements or observations that are made. It is represented as a linear relationship between the observations at the current time and the state at the current time, with an observation matrix. The general formula for the observation equation is:
\[ y(t) = C \cdot x(t) + v(t) \]
where:
\begin{itemize}
  \item $y(t)$ is the observation at time $t$.
  \item $C$ is the observation matrix.
  \item $v(t)$ is the observation noise.
\end{itemize}

In summary, the state space model uses a state equation to describe the evolution of the hidden state of the system and an observation equation to relate the state to the observations. The parameters $A$, $B$, $C$, and the noise matrices $w(t)$ and $v(t)$ are estimated to fit the model to the observed data.

Additionally, for the implementation of the models in the application, we utilized the auto.arima() and StructTS() functions from the forecast and stats libraries in R. These functions are essential for the analysis and forecasting of time series.

The auto.arima() function was employed to automatically select the optimal ARIMA model for the time series of mining production in Peru. On the other hand, the StructTS() function was used to fit a State Space model to the time series.

These functions enabled us to conduct detailed analysis and generate accurate forecasts for mining production. By selecting the appropriate ARIMA model and fitting a State Space model, we obtained valuable information for decision-making in the mining sector.

\subsection{Model Validation and Performance Evaluation}
In this stage, we used residual analysis, including the Ljung-Box test and Shapiro‒Wilk Normality test \cite{dabral2017modelling}, to assess the quality of fit of the ARIMA and State Space models \cite{gerlach1999diagnostics}. Additionally, we generated bootstrap data to evaluate the performance of the application and obtain uncertainty estimates for mining production predictions \cite{ledesma2008introduccion}. These techniques allowed us to assess the accuracy and reliability of the models and the developed application.

\subsection{Application Development}
The MineAnalytica application was developed in Shiny/
RStudio, making use of the following R libraries and packages:
Shiny, Shinydashboard, Leaflet, sf, Readxl, Plotly, Lubridate, Openxlsx, DT, Dplyr, Tidyverse, Magick, Scales, Forecast, Shinyjs, and FontAwesome. These libraries and packages provided specific functionalities for data visualization, interactive chart generation, and prediction within the application.

\section{Results}

In this section, we will present the main results obtained from the analysis and processing of mining production data in Peru, as well as the predictions generated by the time series models implemented in the MineAnalytica application.

\subsection{Prediction}\label{Prediction}
The prediction is based on the analysis and processing of the collected mineral extraction data using the ARIMA model and other R packages such as "forecast". This approach allows for fitting an optimal model to the historical data and generating estimations for future periods. The predictions based on minerals and departments with limited historical data only forecast the next 3 months, while the total annual prediction extends up to 5 years as the data span from 1980 to 2022.

To evaluate the performance of the model, the Ljung-Box test was applied. The test result for (P value = 0.5986) \textgreater\ ($\alpha$ = 0.05) indicates a good performance indicator for the variable (COPPER). Following this, the predictive model for the other metals for the next 5 years is presented. For the year 2027, the following quantities are estimated as predictions for the total annual extraction:

\begin{table}[htbp]
\centering
\caption{Annual extraction prediction for the year 2027}
\begin{tabular}{cc}
\hline
\textbf{Mineral} & \textbf{Annual Extraction} \\
\hline
Copper & 2,694,957 MT \\
Gold & 72,817.47 Fine KG \\
Zinc & 1,369,649 MT \\
Silver & 3,083,036 MT \\
Lead & 255,443 MT \\
Iron & 15,776,609 MT \\
Tin & 29,542 MT \\
Molybdenum & 35,044.66 MT \\
Cadmium & 724 MT\\
\hline
\end{tabular}
\end{table}

These predictions are based on historical data and provide valuable information for strategic decision-making and resource management in the mining sector.

\subsubsection{Annual Mineral Extraction Prediction}
Using historical data of mineral extraction, the ARIMA model can identify patterns and trends in production and generate future projections. These predictions are backed by statistical analysis and allow users to gain more accurate and informed insight into future mineral production in Peru's departments.

\begin{figure}[h]
    \centering
    \includegraphics[scale=0.195]{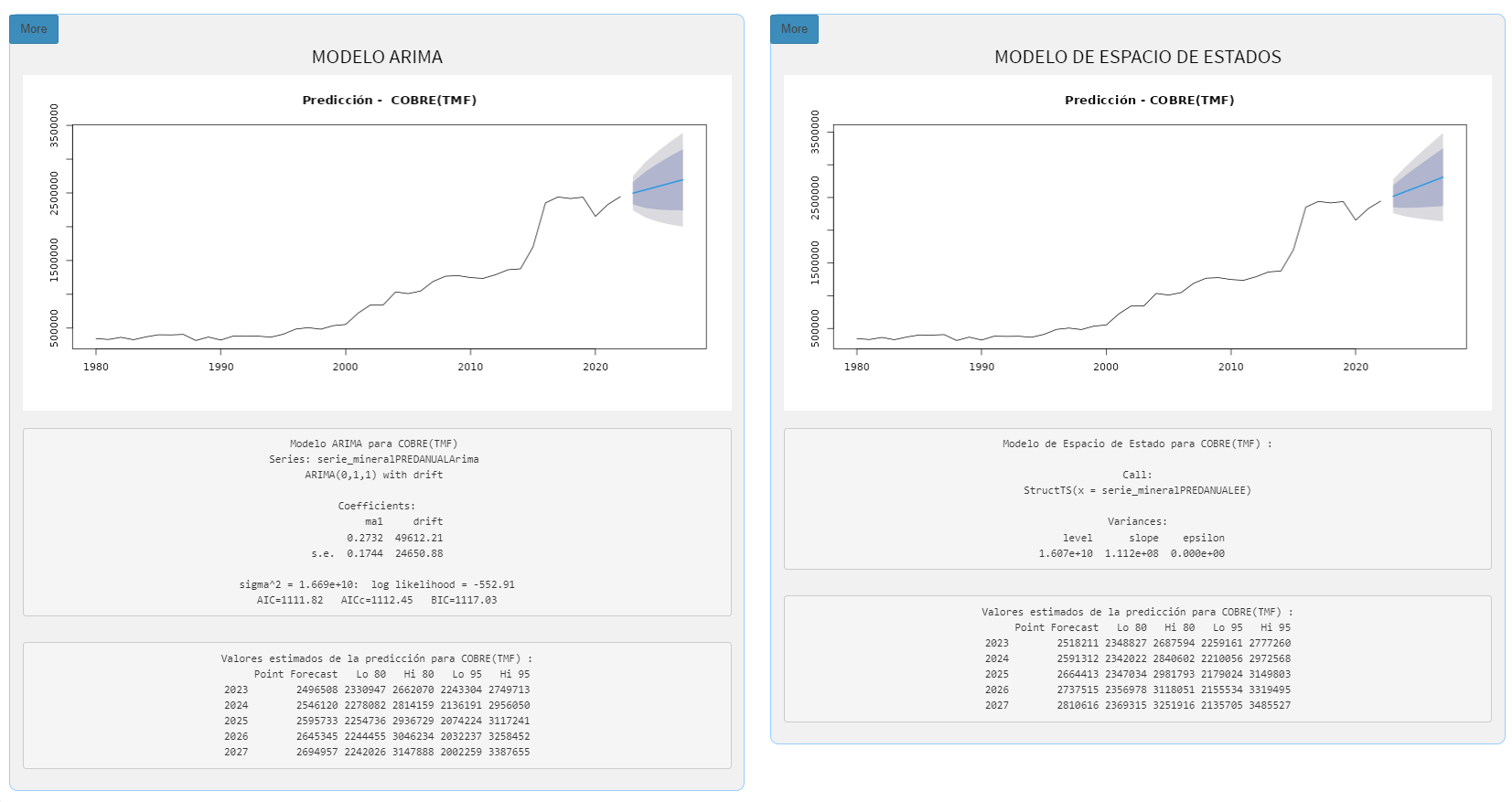}
    \caption{Annual extraction forecast (5 years)}
    \end{figure}

\subsubsection{Mineral-based Prediction}
These predictions enable users to anticipate the amount of mineral expected to be extracted in a specific period and make strategic decisions accordingly. Furthermore, the predictions can also be useful for demand planning, resource management, and decision-making in the mining sector.

\begin{figure}[h]
    \centering
    \includegraphics[scale=0.245]{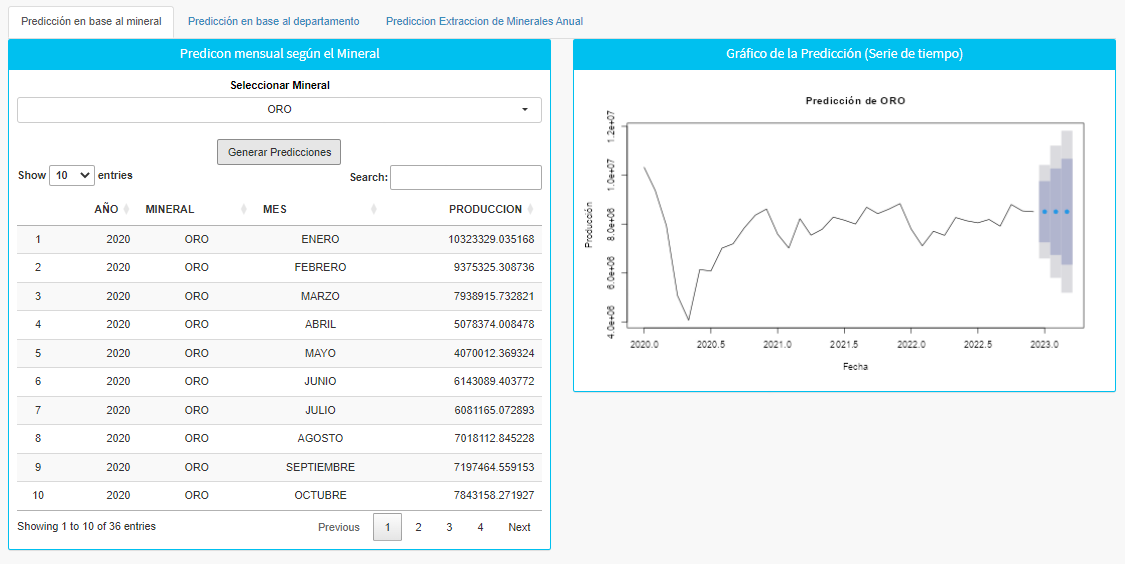}
    \caption{Mineral forecasting (3 months)}
    \end{figure}

\subsubsection{Department-based Prediction}
Using historical data of mineral extraction, the ARIMA model can identify patterns and trends in production and generate future projections. These predictions are backed by statistical analysis and allow users to gain more accurate and informed insight into future mineral production in Peru's departments.

\begin{figure}[h]
    \centering
    \includegraphics[scale=0.245]{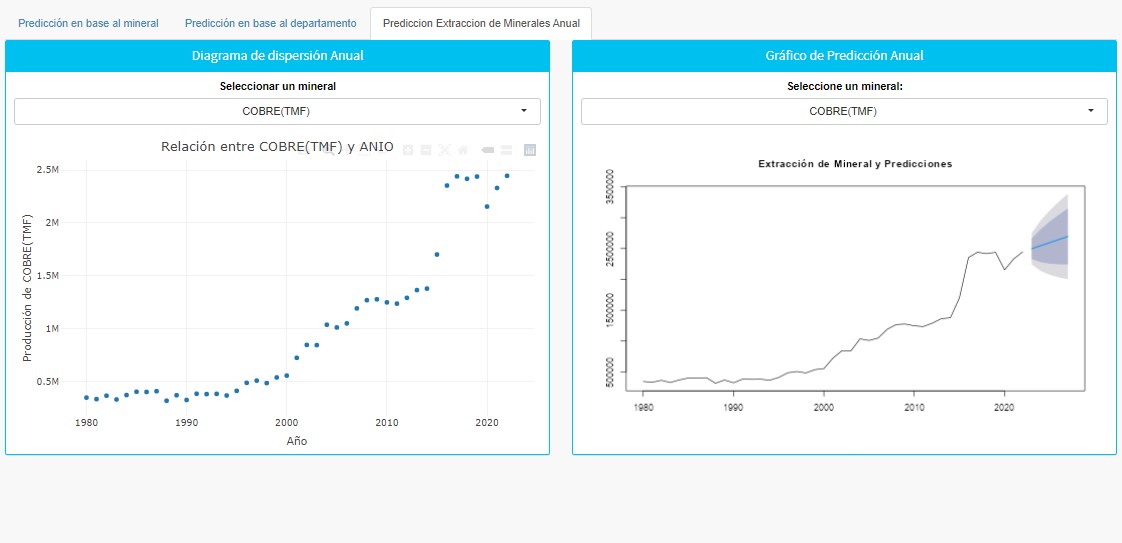}
    \caption{Forecast based on department (3 months)}
    \end{figure}

It is important to note that predictions are subject to uncertainty, and results may vary based on different factors such as changes in demand, price fluctuations, or market conditions. However, the application of prediction techniques based on historical data can provide valuable guidance for decision-making and planning in the mining sector regarding specific mineral extraction in Peru.

\subsection{Application}
\subsubsection{Data Processing}
Various techniques were employed for data processing, including the K-Nearest Neighbors (KNN) algorithm and the Extract, Transform, Load (ETL) process \cite{lozano2018automatizacion}. The ETL process facilitated the extraction, transformation, and loading of the data, ensuring its quality and preparation for subsequent analysis. The KNN algorithm was used in the data transformation process.

%\begin{figure}[h]
%    \centering
%    \includegraphics[scale=0.245]{DATOS1.jpg}
%    \caption{Data display}
%    \end{figure}

\subsubsection{Graphs}
Graphical representations are an effective way to understand statistics visually, and they are an essential component of this application.

Bar graphs and pie charts are widely used visual tools in data analysis. Bar graphs allow for clear and concise comparison of categories or variables, while pie charts highlight the relative distribution of data across categories, showing the proportion of each category in the dataset. Both representations are effective and easy to understand, providing a quick and clear visualization of the information.

\begin{figure}[h]
    \centering
    \includegraphics[scale=0.215]{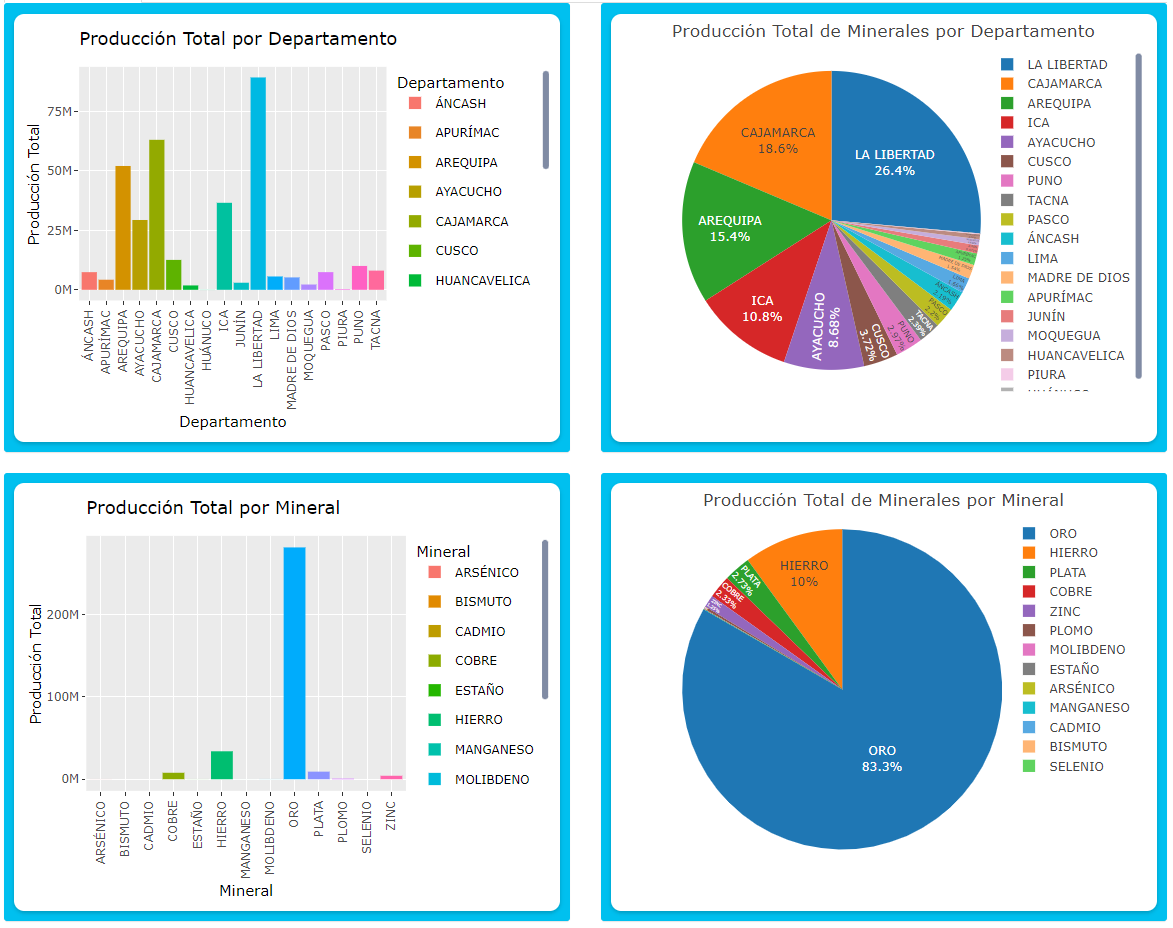}
    \caption{Pie charts}
    \end{figure}

Frequency polygon graphs: Frequency polygons are a useful tool for visualizing and analyzing the frequency distribution of continuous variables. They provide a clear graphical representation of patterns and trends in the data, enabling a deeper understanding of the distribution and variability of the variable.

\begin{figure}[h]
    \centering
    \includegraphics[scale=0.235]{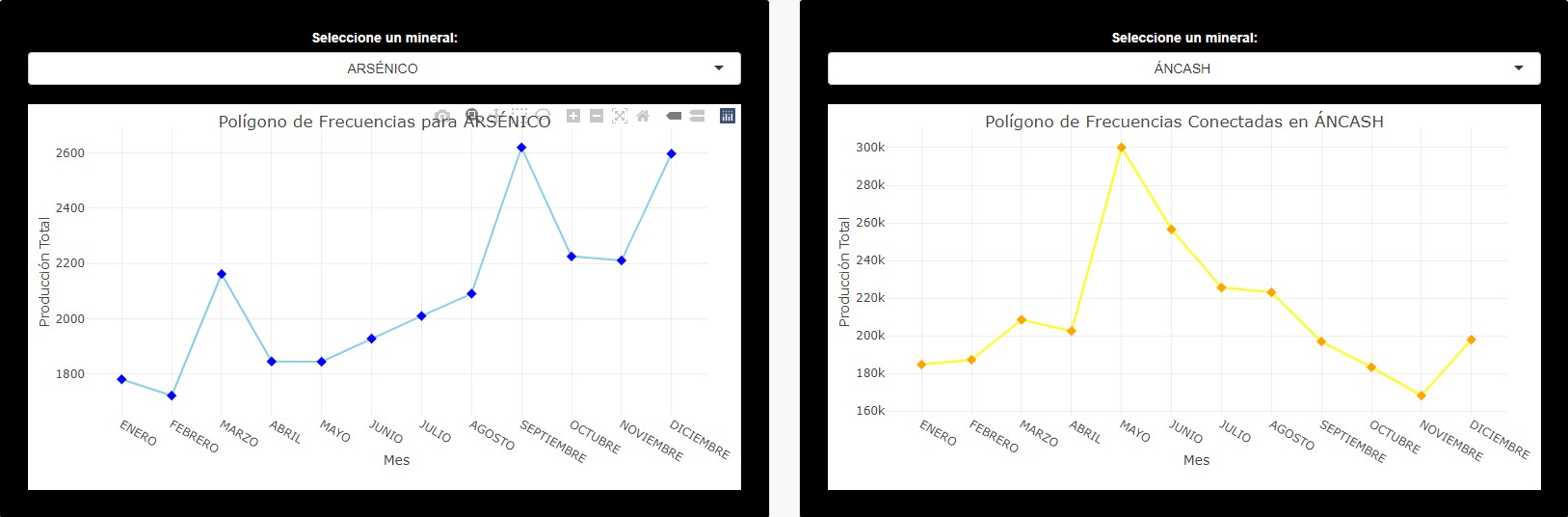}
    \caption{Frequency polygon plots}
    \end{figure}

\subsubsection{Map}
By using this application, users can intuitively and visually explore the geography of Peru and obtain relevant data about each selected department. The application utilizes data visualization techniques and graphs to present information in a clear and understandable manner.

\begin{figure}[h]
    \centering
    \includegraphics[scale=0.205]{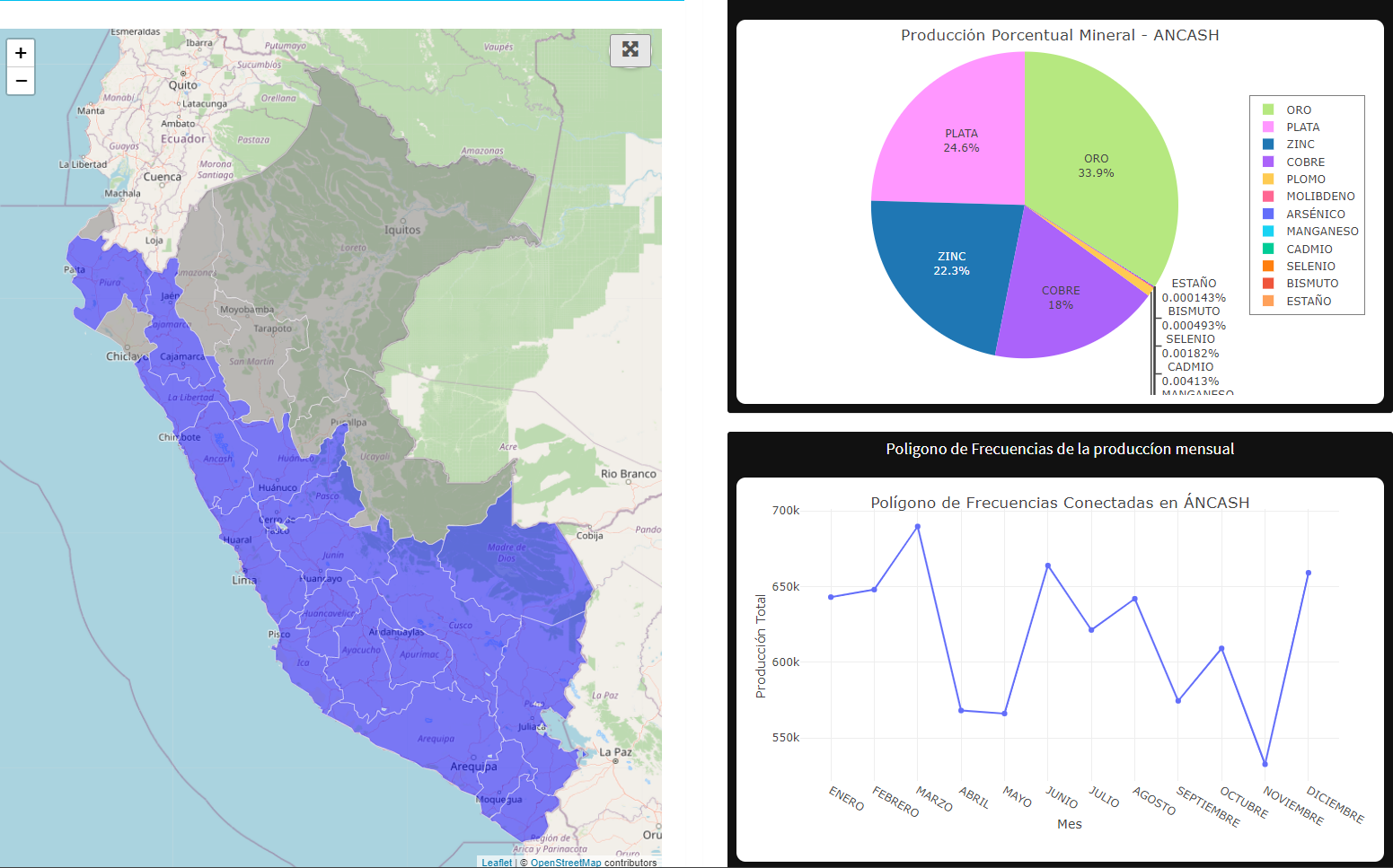}
    \caption{Peru Map}
    \end{figure}

Clicking on a specific department on the map will display a popup window or section in the application showing relevant statistics. This provides a valuable tool for exploring and understanding the diversity and particularities of each region in the context of mining in Peru.

For those who wish to interact with the application, it can be accessed through the following link: url = https://yhackpaco.shinyapps.io/MineAnalytica/

\section{Discussion}

The continuous growth of the mining industry is driven by long-term global trends. Contrary to initial predictions, in countries belonging to the Organization for Economic Cooperation and Development (OECD) \cite{oecd}, the demand for minerals has not decreased as they reach higher levels of national economic development \cite{bebbington2009institutional}. Our predictions confirm this trend.

In the field of mining safety and health, there is evidence of fatal accidents in this activity. Within the period from 2000 to 2006, it was found that thirty-three companies out of a total of eighty-four accounted for 80\% of fatal accidents. Among these companies, six are considered large (18\%), twenty-five are medium-sized (76\%), and three are small (6\%) \cite{lira2007panorama}. This should indicate a decrease in mining production in small companies, but with the graphical analysis, we can see that it is increasing.

The inherent variability makes it difficult to predict the price of these metals, as it is subject to wide and unpredictable movements that can have lasting effects. This is evident in the notable alternation between growth percentages from one year to another. A recent example of this is the drastic change that occurred between 2008 and 2009, with a growth of 37\%, followed by several periods of negative variation starting from 2011 \cite{sthioul2015aportes}. Although the idea that prediction is very uncertain is shared, statistically, we can predict the amount of mineral extraction for the mentioned metals in 5 years.

The production of silver based on data from 1986 to 2013 predicts a decrease in production by 2025 \cite{acosta2014plata}. In contrast, with our model, we predict an increase in production for that year and onward.

These discussions provide an opportunity to analyze and reflect on the obtained results, as well as to highlight the practical and theoretical implications of this project in the context of the Peruvian mining industry.

\section{Conclusions}
Statistical analysis applied to the Peruvian mining industry provides valuable tools for understanding trends, patterns, and factors that influence mining production. This allows for the identification of improvement opportunities and optimization of operational processes.

The ARIMA model is an effective statistical methodology for analyzing and predicting time series data, such as mineral extraction in the Peruvian mining industry. It enables the identification of patterns and trends in historical data and generates predictions to support decision-making.

The application of the ARIMA methodology in the Peruvian mining industry can help anticipate future scenarios and support strategic decision-making. This can contribute to more efficient management of mineral resources and sustainable development of the industry.

Furthermore, to enhance the user experience, an interactive map of Peru was incorporated into the application. This allows for an intuitive visualization of the geographic distribution of mining regions and facilitates understanding of the importance and impact of the mining industry in different areas of the country. The inclusion of this interactive map enriches the user experience by providing a more comprehensive and engaging visual representation of mining data.

In conclusion, statistical analysis and the ARIMA model are essential tools for understanding and optimizing the Peruvian mining industry. They enable the identification of patterns, prediction of trends, and informed and strategic decision-making. By utilizing these tools, the responsible and sustainable development of the mining industry in Peru can be promoted.

\section*{Acknowledgements}
We would like to express our sincere gratitude to the Universidad Nacional del Altiplano, in particular to the School of Statistical Engineering and infomatics, for giving us the invaluable opportunity to grow as professionals in their classrooms. Its commitment to academic excellence and the integral development of students has been fundamental in our formation.

We would like to extend our thanks to all our esteemed teachers, who have given us unconditional support from the very beginning. Their dedication and guidance have been a constant source of inspiration in our educational journey. In particular, we would like to highlight Professor Fred Torrez Cruz, whose motivation and tireless support pushed us to overcome the challenges and successfully complete this project.

Likewise, we cannot forget to mention our classmates, who have been an integral part of our educational journey. Their perspectives, collaboration and mutual support have enriched our training and propelled us to joint achievements. Through interaction and the exchange of ideas, we have grown as professionals and forged lasting friendships.

Finally, we would like to express our deep appreciation to our families and loved ones, who have given us their unconditional support throughout this process. Their encouragement and understanding have given us the strength to face the challenges and pursue our academic dreams.

\printbibliography

@article{newbold1983arima,
  title={ARIMA model building and the time series analysis approach to forecasting},
  author={Newbold, Paul},
  journal={Journal of forecasting},
  volume={2},
  number={1},
  pages={23--35},
  year={1983},
  publisher={Wiley Online Library}
}

@book{cryer1986time,
  title={Time series analysis},
  author={Cryer, Jonathan D},
  volume={286},
  year={1986},
  publisher={Duxbury Press Boston}
}

@article{ledesma2008introduccion,
  title={Introducción al Bootstrap. Desarrollo de un ejemplo acompañado de software de aplicación},
  author={Ledesma, Rubén},
  journal={Tutorials in quantitative methods for psychology},
  volume={4},
  number={2},
  pages={51--60},
  year={2008}
}

@article{Sanchez2017Analisis,
  title={Análisis estadístico para la optimización de la producción minera en el Perú},
  author={Sanchez, Yelitza Josefina and Talavera Pereira, Rosalba},
  journal={Revista de minería y economía},
  year={2017},
  pages={56--72}
}

@article{zarate2020industria,
  title={La industria extractiva en América Latina, su incidencia y los conflictos socioambientales derivados del sector minero e hidrocarburos},
  author={Zárate, Ruth and Vélez, Claudia L and Caballero, José A},
  journal={Revista ESPACIOS. ISSN},
  volume={798},
  pages={1015},
  year={2020}
}

@article{cardenas2016vulnerabilidad,
  title={Vulnerabilidad social y la minería en el Perú: un análisis comparativo},
  author={Cárdenas, Martha Jhiannina and Saraiva, María},
  journal={Revista de Ciencia Política y Gobierno},
  volume={3},
  number={6},
  pages={231--249},
  year={2016}
}

@online{ingemmet,
  title={Instituto Geológico, Minero y Metalúrgico (INGEMMET)},
  url={https://www.gob.pe/ingemmet},
  urldate={2023-06-12}
}

@online{bcrp,
  title={Estadísticas económicas - Banco Central de Reserva del Perú},
  url={https://estadisticas.bcrp.gob.pe/estadisticas/series/mensuales/resultados/PN37543BM/html},
  urldate={2023-06-12}
}

@article{lalama2017,
  title={Capitalismo social: un vistazo a resultados macroeconómicos de Ecuador, Perú y Colombia},
  journal={Revista de Ciencia Sociales},
  author={Lalama, R. A. and Bravo, A.},
  volume={25},
  pages={12--24},
  year={2017}
}

@online{snmpe,
  title={Sociedad Nacional de Minería, Petróleo y Energía (SNMPE)},
  url={https://www.snmpe.org.pe/},
  urldate={2023-06-12}
}

@online{minem,
  title={Ministerio de Energía y Minas},
  url={https://www.gob.pe/minem},
  urldate={2023-06-12}
}

@article{szablowski2002mining,
  title={Mining, displacement and the World Bank: A case analysis of compañia minera antamina's operations in Peru},
  author={Szablowski, David},
  journal={Journal of Business Ethics},
  volume={39},
  pages={247--273},
  year={2002},
  publisher={Springer}
}

@misc{forecast_page,
  title={RStudio Forecast Function},
  howpublished={\url{https://rstudio-pubs-static.s3.amazonaws.com/556459_38e6ed9ddfee4a27a0fc5814d12cd416.html}},
  note={Accedido el 23 de junio de 2023},
}

@article{garcia2018tratamiento,
  title={Tratamiento de datos para el análisis estadístico: Métodos y técnicas},
  author={García, A. and López, M. and Martínez, P.},
  journal={Revista de Estadística Aplicada},
  year={2018},
}

@article{lozano2018automatizacion,
  title={Automatización en la recolección, tratamiento y envío de información estadística médico-asistencial a la Superintendencia Nacional de Salud (SUSALUD) basado en procesos de ETL y RPA para la clínica Adventista Ana Stahl},
  author={Lozano, Lizana and Mical, Rosi},
  year={2018},
  publisher={Universidad Peruana Unión}
}

@misc{oecd,
    title={About the OECD},
    howpublished={\url{https://www.oecd.org/}},
    note={Accedido el 23 de junio de 2023},
}

@article{bebbington2009institutional,
  title={Institutional challenges for mining and sustainability in Peru},
  author={Bebbington, Anthony J and Bury, Jeffrey T},
  journal={Proceedings of the National Academy of Sciences},
  volume={106},
  number={41},
  pages={17296--17301},
  year={2009},
  publisher={National Academy of Sciences}
}

@book{lira2007panorama,
  title={Panorama de la Minería en el Perú},
  author={Lira, Alfredo Dammert and Aristondo, Fiorella Molinelli},
  year={2007},
  publisher={Osinergmin}
}

@article{sthioul2015aportes,
  title={Aportes de la minería a Chile y Perú: Interacción con la sociedad},
  author={Sthioul Ortíz, Alberto Enrique},
  year={2015},
  publisher={Universidad de Chile}
}

@article{acosta2014plata,
  title={Plata: antecedentes y proyecciones hasta el 2025},
  author={Acosta Ale, Jorge Gilberto and Santisteban Angeldonis, Alexander and Huanacuni Mamani, Dina and Valencia Muñoz, Michael Melitón and Villarreal Jaramillo, Eder},
  year={2014},
  publisher={Horizonte Minero}
}

@article{dabral2017modelling,
  title={Modelling and forecasting of rainfall time series using SARIMA},
  author={Dabral, PP and Murry, Mharhoni Z},
  journal={Environmental Processes},
  volume={4},
  number={2},
  pages={399--419},
  year={2017},
  publisher={Springer}
}

@article{gerlach1999diagnostics,
  title={Diagnostics for time series analysis},
  author={Gerlach, Richard and Carter, Chris and Kohn, Robert},
  journal={Journal of Time Series Analysis},
  volume={20},
  number={3},
  pages={309--330},
  year={1999},
  publisher={Wiley Online Library}
}

@inproceedings{chu2016data,
  title={Data cleaning: Overview and emerging challenges},
  author={Chu, Xu and Ilyas, Ihab F and Krishnan, Sanjay and Wang, Jiannan},
  booktitle={Proceedings of the 2016 international conference on management of data},
  pages={2201--2206},
  year={2016}
}

@book{aoki2013state,
  title={State space modeling of time series},
  author={Aoki, Masanao},
  year={2013},
  publisher={Springer Science \& Business Media}
}
\end{document}